\documentclass[12pt]{article}
\usepackage{a4,amsmath,epsfig,amssymb}

\begin{document}

\begin{titlepage}
\begin{flushright}
PCCF RI 0417 \\
 ${\rm ECT}^{*}$-04-22\\
\end{flushright}
\renewcommand{\thefootnote}{\fnsymbol{footnote}}
\vspace{1.0em}

\begin{center}
{\bf \LARGE{ $\boldsymbol{\Lambda_b}$ Decays into  $\boldsymbol{\Lambda}$-Vector}}
\end{center}
\vspace{1.0em}
\begin{center}
\begin{large}
Z.J. Ajaltouni$^{1}$\footnote{ziad@clermont.in2p3.fr}, 
E. Conte$^{2}$\footnote{conte@clermont.in2p3.fr},
O. Leitner$^{3}$\footnote{leitner@ect.it} \\
\end{large}
\vspace{2.3em}
$^{1,2}$ Laboratoire de Physique Corpusculaire de Clermont-Ferrand \\
IN2P3/CNRS Universit\'e Blaise Pascal \\
F-63177 Aubi\`ere Cedex France \\ 
$^3$ ${\rm{ECT}}^{*}$, Strada delle Tabarelle, 286, 
38050 Villazzano (Trento), Italy \\
\end{center}
\vspace{8.0em}
\begin{abstract}
\vspace{1.0em}
A complete study of the angular distributions of the processes, 
$\Lambda_b \to {\Lambda} V(1^-)$, with $\Lambda \to p {\pi}^-$ and $V (J/{\Psi}) 
\to {\ell}^+ {\ell}^-$ or $V ({\rho}^0,\omega) \to {\pi}^+ {\pi}^-,$  is performed. 
 Emphasis is put on the initial $\Lambda_b$ polarization produced in the 
proton-proton collisions. The polarization density-matrices as well as 
angular distributions  are derived and help to construct  T-odd observables 
which allow us to perform tests of both Time-Reversal and $CP$ violation. 
\vskip 0.3cm
\end{abstract}
\vspace{16.em}
PACS Numbers:  11.30.Er, 12.39.-x, 13.30.-a, 14.20.-c, 14.20.Mr
\newline
Keywords: Helicity, baryon decay, polarization
%
\end{titlepage}
\newpage
%
\section{Introduction}\label{section1}
%
With the advent of $B$-factories at the proton-proton colliders, huge 
statistics of beauty hadrons are expected to be produced. This will allow 
a thorough study of $CP$ violation processes with  $B$ mesons. Moreover, 
some specific phenomena related to either $b$-quark physics or $CP$ violation 
can be performed to put limits on the validity of the Standard Model (SM). 
One of these processes concerns the validity of the Time Reversal (TR) 
symmetry. A promising method to look for TR violation is the  three body 
$\Lambda_b$ decay~\cite{Bensalem:2002pz, Chen:2002ja} as it was 
initiated with the hyperons long time ago by R. Gatto~\cite{Gatto:1958}. 

T-odd operator is derived from Time Reversal and it keeps the initial and 
final states  unchanged. It is well known that the time reversing state of 
a decay like $\Lambda \to p {\pi}^-$ or $\beta$ nucleon decay cannot be 
realized in the physical world, thus we must be contented with the following 
transformations which are the main ingredients of TR operator:
\begin{eqnarray*}
\vec r   \xrightarrow{T} \vec r\ , \;\;   \vec p   \xrightarrow{T}  -{\vec p}\ , \;\;  
\vec \ell  \xrightarrow{T}  -{\vec \ell}\ ,  \;\;  \vec s  \xrightarrow{T}  -{\vec s}\ , 
\end{eqnarray*}
where $\vec \ell$ and $\vec s$ are respectively the angular momentum and 
the spin of any particle with momentum $\vec p$. Consequently the helicity 
of the particle defined by $\lambda = {\vec s}\cdot{\vec p}/p$
remains unchanged by TR transformations.

In the past, it was pointed out by many authors~\cite{Valencia:1994zi} 
the importance to look for T-odd effects in the hyperon decays like 
$\Lambda, \Sigma$ and $\Xi$, as being a consequence of both  $CPT$ theorem 
and $CP$ violation in weak $|{\Delta}S| = 1$ decays. As far as beauty hadrons 
${\Lambda}_b, {\Sigma}_b$ and ${\Xi}_b$ are concerned,  because of their 
numerous decay channels and the strength of  $CP$ violation in the $b$-quark 
sector,  opportunities to find T-odd observables will increase and interesting 
tests of both the SM  and models beyond the SM can be performed successfully. 
Due to the initial polarization of the $\Lambda_b$ baryon, T-odd observables 
can be constructed from the decay products such as $\vec {v_1} \cdot {(\vec {v_2} 
\times \vec {v_3})}$ where $\vec {v_i}$ is either the spin or the momentum of 
the particle $i$.  These observables change sign under TR transformations 
and a non-vanishing  mean value of their distribution could be a sign of 
TR violation. 

This paper is  devoted to a  study and  simulations of $\Lambda_b$ decays into 
$\Lambda {\ell}^+ {\ell}^-$ and  $\Lambda h^+ h^-$. Final leptons, $\ell = e, \mu$, 
or final hadron, $h=\pi$, can originate from intermediate resonances which quantum 
numbers are those of a vector meson $1^-$  like  $J/\psi, {\rho}^0$ and $\omega$. 
The reminder of this paper is organized as it follows. In section 2, we present 
an analysis of both the intermediate states and the final particles in some 
appropriate frames, the  helicity frames. By stressing on the importance of 
the polarizations of the initial $\Lambda_b$ as well as the intermediate 
resonances, calculations based on the helicity formalism are performed and 
take into account the spin properties of the final decay products. Dynamical 
assumption is made through the factorization framework applied in baryon 
decays in section 3.  The following section is devoted to results and discussions 
for angular distributions and polarization density matrices. Finally, in the 
last section, we  draw some conclusions.
%
\section{$\boldsymbol{\Lambda_b}$ decay analysis}\label{section2}
%
In the collisions, $p p \rightarrow  {\Lambda_b} + X$, the $\Lambda_b$ is 
produced with a  transverse polarization in a similar way than the 
ordinary hyperons. Its longitudinal polarization is suppressed because 
of parity conservation in strong interactions. Let us define, 
${\vec N}_P$, the vector normal to the production plane by: 
\begin{eqnarray}
  {\vec N}_P =  \frac {\vec{p}_1 \times \vec{p}_b} {|{\vec{p}_1 \times \vec{p}_b}|}\ , 
\end{eqnarray} 
where $\vec{p}_1$ and $\vec{p}_b$ are  the vector-momenta of one incident 
proton beam and $\Lambda_b$, respectively. The mean value of the $\Lambda_b$
spin along ${\vec N}_P$ is the $\Lambda_b$ transverse polarization  usually  
greater than $20\%$~\cite{Leader:2001}.
  
Let $({\Lambda_b}xyz)$ be the rest frame (see Fig.~1) of the $\Lambda_b$ 
particle. The quantization axis $({\Lambda_b}z)$ is chosen to be parallel 
to ${\vec N}_P$. The other orthogonal axis $({\Lambda_b}x)$ and $({\Lambda_b}y)$ 
are chosen arbitrarily in the production plane. In our analysis, 
the $({\Lambda_b}x)$ axis is taken parallel to the momentum $\vec {p}_1$. 
The spin projection, $M_i$, of the $\Lambda_b$ along the  transverse axis 
$({\Lambda_b}z)$ takes  the values   $\pm {1/2}$. The  polarization density 
matrix\footnote{The polarization density matrix elements (PDM), 
${\rho}^{\Lambda_b}_{ij}$, do not need to be exactly known since the initial, 
$\Lambda_b,({\cal P}_{\Lambda_b}=\rho_{11}^{\Lambda_b} - \rho_{22}^{\Lambda_b})$ 
polarization~\cite{Leader:2001} is only required in our analysis.}, 
${\rho}^{\Lambda_b}$, of the $\Lambda_b$ is a $(2 \times 2)$ hermitian matrix. 
Its elements\footnote{Note as well that ${\rho}_{ij}^{\Lambda_b} = ({\rho}_{ji}^{\Lambda_b})^{*}$.},      
  ${\rho}_{ii}^{\Lambda_b}$,  are real and $\sum_{i=1}^{2} {\rho}_{ii}^{\Lambda_b}= 1$.  
The  probability of having $\Lambda_b$  produced with $M_i = \pm{1/2}$ is 
given by ${\rho}_{11}^{\Lambda_b}$ and ${\rho}_{22}^{\Lambda_b}$, respectively. 
Finally, the initial $\Lambda_b$ polarization, ${\cal P}_{\Lambda_b}$, 
is  given by $\langle \vec{S}_{\Lambda_b} 
\cdot \vec{N}_P \rangle = {\cal P}_{\Lambda_b}
={\rho}_{11}^{\Lambda_b} - {\rho}_{22}^{\Lambda_b}.$

The decay amplitude, $A_0{(M_i)}$, for ${\Lambda_b}{(M_i)}  \to  
{\Lambda}{(\lambda_1)} V{(\lambda_2)}$
is obtained by applying the Wigner-Eckart theorem to the $S$-matrix
element in the framework of the Jacob-Wick helicity formalism~\cite{Jackson:1965}:
\begin{eqnarray}
 A_0{(M_i)} = \langle 1/2,M_i|S^{(0)}|p,\theta,\phi;\lambda_1,\lambda_2 \rangle 
= \mathcal{M}_{\Lambda_b}{(\lambda_1,\lambda_2)} D_{M_i M_f}^{{1/2} \star}{(\phi,\theta,0)}\ , 
\end{eqnarray}
where $\vec p = (p,\theta,\phi)$ is the vector-momentum of the hyperon 
$\Lambda$ in the $\Lambda_b$ frame (Fig.~1). $\lambda_1$ and $\lambda_2$ are the 
respective helicities of $\Lambda$  and $V$ with the possible values $\lambda_1=\pm{1/2}$ 
and  $\lambda_2 =-1,0,+1$. The momentum projection along the 
$({\Delta})$ axis (parallel to $\vec{p}$) is given by $M_f =\lambda_1-\lambda_2=\pm{1/2}$.
The $M_f$ values  constrain those of $\lambda_1$ and $\lambda_2$ since, among six 
combinations, only  four are physical. If $M_f= +1/2$ then $(\lambda_1,\lambda_2)=(1/2,0)$ or $(-1/2,-1)$.
If $M_f= -1/2$ then $(\lambda_1,\lambda_2)=(1/2,1)$ or $(-1/2,0)$.
The hadronic matrix element, $\mathcal{M}_{\Lambda_b}{(\lambda_1,\lambda_2)}$, 
contains  all the  decay dynamics. Finally, the Wigner matrix element,
\begin{eqnarray}
 D_{M_i M_f}^j{(\phi,\theta,0)} = d_{M_i M_f}^j{(\theta)} {\exp{(-i{M_i}{\phi})}}\ ,
\end{eqnarray} 
is expressed according to the Jackson convention~\cite{Jackson:1965}. 

In case of two intermediate resonances such as those described in the next section, 
the $\Lambda_b$-decay plane is defined by the momenta 
of the $\Lambda$ and leptons (or hadrons). This decay plane does not 
coincide with that one defined by the momenta of the $J/\Psi$, proton and pion.
%
\subsection{Decay of the intermediate resonances}
%
By performing appropriate rotations and Lorentz boosts, we can study the decay 
of each resonance in its own helicity frame (see Fig.~1) such that the 
quantization axis is parallel to the resonance momentum in the $\Lambda_b$ 
frame i.e. $\overrightarrow{O_1z_1} || \vec{p}_{\Lambda}$ and 
$\overrightarrow{O_2z_2} || \vec{p}_{V}=-\vec {p}_{\Lambda}$.  For the decays 
${\Lambda}{(\lambda_1)} \to  P{(\lambda_3)} {\pi}^-{(\lambda_4)}$  and 
$V{(\lambda_2)} \to  {\ell}^-{(\lambda_5)} {\ell}^+{(\lambda_6)}$  
or $V{(\lambda_2)} \to h^-{(\lambda_5)} h^+{(\lambda_6)},$ the respective 
helicities of the final particles are $(\lambda_3,\lambda_4)=(\pm{1/2},0)$ 
and $(\lambda_5,\lambda_6)=(\pm{1/2},\pm{1/2})$ in case of 
leptons or $(\lambda_5,\lambda_6)=(0,0)$ in case of  $0^-$ mesons.

In the $\Lambda$ helicity frame, the projection of the total angular momentum, 
$m_i$, along the proton momentum, $\vec{p}_P$, is given by $m_1=\lambda_3-\lambda_4 =\pm{1/2}$. 
In the vector meson helicity frame, this projection is equal to $m_2=\lambda_5-\lambda_6= -1,0,+1$ 
if leptons and $m_2=0$ if  $\pi$. The decay amplitude, $A_i(\lambda_i)$, of 
each resonance can be written similarly as in Eq.~(2), requiring only that the 
kinematics of its decay products are fixed. We obtain,
\begin{eqnarray}
A_1(\lambda_1) =\langle \lambda_1, m_1|S^{(1)}|p_1,{\theta}_1,{\phi}_1;\lambda_3,\lambda_4 \rangle =  
 \mathcal{M}_{\Lambda}{(\lambda_3,\lambda_4)} D^{{1/2} \star}_{\lambda_1 m_1}{(\phi_1,\theta_1,0)}\ , \nonumber \\
A_2(\lambda_2) =\langle  \lambda_2, m_2|S^{(2)}|p_2,{\theta}_2,{\phi}_2;\lambda_5,\lambda_6 \rangle  =  
 \mathcal{M}_{V}{(\lambda_5,\lambda_6)} D^{1 \star}_{\lambda_2 m_2}{(\phi_2,\theta_2,0)}\ , 
\end{eqnarray}
where  $\theta_1$ and  $\phi_1$ are respectively the polar and azimuthal angles 
of the proton momentum in the $\Lambda$ rest frame while $\theta_2$ and $\phi_2$ 
are those of ${\ell}^- {(h^-)}$ in the $V$ rest frame.
\subsection{Analytical form of the decay probability}
%
The general decay amplitude\footnote{We assume that the three decay amplitudes, 
$A_0(M_i), A_1(\lambda_1)$ and $A_2(\lambda_2)$ are independent so that the 
general amplitude, ${\cal A}_I$, is given by the product of the three amplitudes, 
$A_i(\lambda_i)$.}, ${\cal A}_I$, for the process\footnote{In a similar way for 
the process $\Lambda_b{(M_i)} \to \Lambda{(\lambda_1)}  V{(\lambda_2)}  \Longrightarrow  P {\pi}^-  \;  
h^+ h^-.$}, $\Lambda_b{(M_i)} \to \Lambda{(\lambda_1)}  V{(\lambda_2)}  \Longrightarrow  P {\pi}^-  \;  
{\ell}^+ {\ell}^-$,  must include all the possible 
intermediate states so that a sum over the helicity states ${(\lambda_1, \lambda_2)}$ 
is performed:
\begin{eqnarray}
 {\cal A}_I = \sum_{\lambda_1,\lambda_2}{A_0{(M_i)} A_1{(\lambda_1)} A_2{(\lambda_2)}}\ .
\end{eqnarray} 
The decay probability, $d\sigma$, depending on the amplitude,  ${\cal A}_I$, takes 
the form,
\begin{eqnarray}
d\sigma  \propto   \sum_{M_i,M^{\prime}_i}{\rho}_{M_i M^{\prime}_i}^{\Lambda_b} {\cal A}_I {\cal A}_I^*\ ,
\end{eqnarray}
where  the polarization density matrix, ${\rho}_{M_i M^{\prime}_i}^{\Lambda_b}$, 
is used to take into account the unknown $\Lambda_b$ spin component, $M_i$. 
Since the helicities of the final particles  are not measured, a summation over 
the helicity values  $\lambda_3, \lambda_4, \lambda_5$ and  $\lambda_6$  is performed 
as well. Finally, the decay probability, $d\sigma$, written in a such way that  only 
the intermediate resonance helicities appear, reads as,
\begin{multline}\label{eq11}
d\sigma  \propto  
\sum_{\lambda_1,\lambda_2,{\lambda}^{\prime}_1,{\lambda}^{\prime}_2} D_{\lambda_1-\lambda_2,
{\lambda}^{\prime}_1 -{\lambda}^{\prime}_2}{(\theta,\phi,0)} {\rho}_{\lambda_1-\lambda_2, 
{\lambda}^{\prime}_1-{\lambda}^{\prime}_2}^{\Lambda_b}{\mathcal{M}_{\Lambda_b}{(\lambda_1,\lambda_2)}}\\
\mathcal{M}_{\Lambda_b}^*{({\lambda}^{\prime}_1,{\lambda}^{\prime}_2)} F^{\Lambda}_{\lambda_1
{\lambda}^{\prime}_1}{(\theta_1,\phi_1)} G^V_{\lambda_2 {\lambda}^{\prime}_2}{(\theta_2,\phi_2)}\ ,
\end{multline}
where $F^{\Lambda}_{\lambda_1 {\lambda}^{\prime}_1}{(\theta_1,\phi_1)}$ and $G^V_{\lambda_2 
{\lambda}^{\prime}_2}{(\theta_2,\phi_2)}$   describing  the decay dynamics of the intermediate 
resonances $\Lambda \to P {\pi}^-$  and $V \to {\ell}^+ {\ell}^-$, respectively, are given in 
Appendix. Because of parity violation in weak hadronic decays, it is assumed that 
$\mathcal{M}_{\Lambda_b}{(\lambda_1,\lambda_2)}$ is not equal to  
$\mathcal{M}_{\Lambda_b}{(-\lambda_1,-\lambda_2)}$. 
%
\section{Factorization procedure}\label{section3}
%
In tree approximation, the effective interaction\footnote{All the terms of the 
effective interaction are extensively defined in literature.}, $\mathcal{H}^{eff}$, 
written as,
\begin{eqnarray}
\mathcal{H}^{eff}=\frac{G_{F}}{\sqrt{2}} V_{qb}V^{\star}_{qs} \sum_{i=1}^2 c_i(m_b) O_i(m_b)\ , 
\end{eqnarray} 
gives the weak following amplitude factorized into,
\begin{eqnarray}
\mathcal{M}_{\Lambda_b}(\Lambda_b \to \Lambda V)= \frac{G_{F}}{\sqrt{2}} V_{qb}V^{\star}_{qs}
f_V E_V \Bigl( c_1 + \frac{c_2}{N_c^{eff}} \Bigr)
\langle \Lambda | \bar{s} \gamma_{\mu} (1-\gamma_{5}) b| \Lambda_b \rangle\ .
\end{eqnarray} 
The CKM matrix elements, $V_{qb}V^{\star}_{qs}$, read as $V_{ub}V^{\star}_{us}$ 
and $V_{cb}V^{\star}_{cs}$, in case of $\Lambda_b \to \Lambda \rho$ and   
$\Lambda_b \to \Lambda J/\Psi$, respectively. The Wilson 
Coefficients, $c_i$, are equal to $c_1=-0.3$ and $c_2=+1.15$.
The hadronic matrix element, $\langle \Lambda | \bar{s} \gamma_{\mu} (1-\gamma_{5}) b| 
\Lambda_b \rangle$, can be derived respecting Lorentz decomposition. Working in HQET, 
it is more convenient to use~\cite{Mannel:1990vg},
\begin{eqnarray}
\langle \Lambda | \bar{s} \gamma_{\mu} (1-\gamma_{5}) b| \Lambda_b \rangle=
\bar{u}_{\Lambda} \Bigl[ \bigl\{ F_1(q^2) + \slash \!\!\! v F_2(q^2) \bigr\} 
\gamma_{\mu} (1-\gamma_5)  \Bigr] u_{\Lambda_b}\ , 
\end{eqnarray}
where the four-velocity of $\Lambda_b$ is $v=P_{\Lambda_b}/M_{\Lambda_b}$. The 
momentum transfer is denoted by  $q=P_{\Lambda_b}-P_{\Lambda}$ and $F_i(q^2)$ 
are the form factors\footnote{We define $F^{\pm}(q^2)=F_1(q^2)\pm F_2(q^2)$, for 
convenience.} involved in the transition $\Lambda_b \to \Lambda$.
The final amplitude\footnote{In Eq.~(11), the factor, $\frac{G_{F}}{\sqrt{2}} V_{qb}V^{\star}_{qs} 
f_V E_V \Bigl( c_1 + \frac{c_2}{N_c^{eff}} \Bigr)$, is not written only for simplicity.}, 
$\mathcal{M}_{\Lambda_b}(\Lambda_b \to \Lambda V)$, depending on the helicity state, 
$(\lambda_{\Lambda},\lambda_V)$, reads as,
\begin{equation}\label{eq18}
   \mathcal{M}_{\Lambda_b} (\Lambda_b \to \Lambda V)  = \left\{\,\,
   \begin{array}{ll}
    {\displaystyle -\frac{P_V}{E_V} \Biggl( \frac{m_{\Lambda_b}+m_{\Lambda}}
{E_{\Lambda}+m_{\Lambda}} F^-(q^2) + 2 F_2(q^2) \Biggl)} \,; & (\lambda_{\Lambda},\lambda_V)=(\frac12,0)\ ,
        \\[0.4cm] 
    {\displaystyle  \frac{1}{\sqrt{2}} \Biggl( \frac{P_{V}}
{E_{\Lambda}+m_{\Lambda}}F^-(q^2)+   F^+(q^2) \Biggl)} \,; & (\lambda_{\Lambda},\lambda_V)=(-\frac12,-1)\ ,
         \\[0.4cm] 
    {\displaystyle \frac{1}{\sqrt{2}}  \Biggl( \frac{P_{V}}
{E_{\Lambda}+m_{\Lambda}}F^-(q^2) - F^+(q^2) \Biggl)   } \,; & (\lambda_{\Lambda},\lambda_V)=(\frac12,1)\ ,
        \\[0.4cm] 
     {\displaystyle  \Biggl(F^+(q^2) + \frac{P^2_{V}}
{E_V (E_V+m_{\Lambda})} F^-(q^2) \Biggl)  } \,; & (\lambda_{\Lambda},\lambda_V)=(-\frac12,0)\ .
       \\ [0.4cm]      
   \end{array}\right.
\end{equation}
The $q^2$ dependence of the transition form factors, $F_i(q^2)$, or $(F^{\pm}(q^2))$,  
resulting from QCD sum rules and HQET~\cite{Huang:1998ek} takes the form as it follows,
\begin{eqnarray}
F_i(q^2)= \frac{F(0)}{1- a\frac{q^2}{m^2_{\Lambda_b}} + b\frac{q^4}{m^4_{\Lambda_b}}}\ ,
\end{eqnarray}
where  the following values $(0.462,-0.0182,-1.76\!\times\!10^{-4})$ and $(-0.077,-0.0685,1.46\!\times\!10^{-3})$ 
correspond to $(F(0),a,b)$ in case of $F_1(q^2)$ and $F_2(q^2)$, respectively. We refer 
to the PDG~\cite{Eidelman:2004wy} for all the numerical values used in our analysis.
%
\section{Results}\label{section5}
%
Departing from the previous relations, physical observables like the helicity 
asymmetry parameter, $\alpha_{As}^{\Lambda_b}$, the polarization density matrices, 
$\rho^{V,\Lambda}$, and the branching ratios, $\mathcal{BR}({\Lambda}_b \to {\Lambda} {\rho}^0)$ 
and $\mathcal{BR}({\Lambda}_b \to {\Lambda} J/\Psi)$, can be evaluated.

Owing to the spin $1/2$ of the $\Lambda_b$, the angular momentum projection 
along the helicity axis (which direction is given by the $\Lambda$ vector-momentum) 
has only two values, $M_i=\pm1/2$, with respective weights 
generally different. The helicity asymmetry parameter, $\alpha_{As}^{\Lambda_b}$, 
defined in Eq.~(14), takes the following values:
\begin{align}
{\alpha}^{\Lambda_b}_{AS}= 98.8\% & \;\;{\rm for}\;\; \Lambda_b \to \Lambda \rho^0\ , \nonumber\\
{\alpha}^{\Lambda_b}_{AS}= 77.7\% & \;\;{\rm for}\;\; \Lambda_b \to \Lambda J/\Psi\ . \nonumber
\end{align}
From these results, the angular momentum projection, $M_i=1/2$, appears to be largely dominant 
in the analyzed decays.

The $\Lambda$-polarization, $\mathcal{P}_{\Lambda}=\rho_{ii}^{\Lambda}-\rho_{jj}^{\Lambda}$,
with $\rho_{ii}^{\Lambda}$ defined in Eq.~(17), can be computed in both decay cases. 
After normalization of $\mathcal{P}_{\Lambda}$, we obtain the values, 
$\mathcal{P}_{\Lambda}=+31\%$, and $\mathcal{P}_{\Lambda}=-9\%$, for $\Lambda_b \to 
\Lambda {\rho}^0$ and $\Lambda_b \to \Lambda J/{\psi}$, respectively.
The other important parameter concerning the spin state of the intermediate resonances is
the density matrix element, $\rho^V_{ij}$, defined in Eq.~(20). Let us focus 
on the matrix element, $\rho^V_{00}$, which is  related to the longitudinal polarization of 
the vector meson V. After calculation, $65.5\%$ and $55.5\%$
are the results for the density matrix element, $\rho^V_{00},$ in case of  $\Lambda_b \to 
\Lambda {\rho}^0$ and $\Lambda_b \to \Lambda J/{\psi}$, respectively.
It is important to notice that these parameters, ${\alpha}^{\Lambda_b}_{AS}$ and 
$\rho^V_{ij},$ (as well as $\rho^{\Lambda}_{ij}$) govern entirely the angular distributions, 
$W_i(\theta_i,\phi_i)$, of the final particles in each resonance frame.

In Fig.~2, are shown the polar angular distributions 
(which do not depend on $\Lambda_b$ initial polarization)
for proton and $l(h)$ coming respectively from $\Lambda$  and V decays. 
In the same figure, the transverse momentum distributions, $P_{\perp}^{P}$  and  
$P_{\perp}^{\pi}$ ($\Lambda$ daughters)  given in the $\Lambda_b$ rest frame, are plotted. 
These distributions look to be discriminant in the investigation of $\Lambda_b$ decay observables.

Finally, the last step is the computation of the branching ratios, $\mathcal{BR}(\Lambda_b \to 
\Lambda {\rho}^0)$ and $\mathcal{BR}(\Lambda_b \to \Lambda J/{\psi})$, which 
requires the calculation of their corresponding widths. The standard 
expression of a decay width, $\Gamma(\Lambda_b \to \Lambda V)$, is given by,
\begin{eqnarray*}
\Gamma(\Lambda_b \to \Lambda V) 
=\frac{E_{\Lambda}+M_{\Lambda}}{M_{\Lambda_b}}\frac{P_V}{16\pi^2} 
\int_{\Omega}|A_0(M_i)|^2 d\Omega\ ,
\end{eqnarray*}
where $E_{\Lambda}$ and $P_{V}$ are respectively the energy and momentum of 
the $\Lambda$ baryon and vector meson in the $\Lambda_b$ rest frame. $\Omega$
corresponds to the decay solid angle. Performing all the calculations and keeping the number of color,
$N_c^{eff}$, to vary between the values 2 and 3 as it is suggested by the factorization hypothesis, we
obtain the following branching ratio results:
\begin{align}
\mathcal{BR}(\Lambda_b \to \Lambda \rho^0) = & (34.0\ , 11.4\ , 3.1) \times 10^{-8}\ ,   \nonumber \\
\mathcal{BR}(\Lambda_b \to \Lambda J/\psi) = & (12.5\ , 4.4\ ,1.2) \times 10^{-4}\ ,  \nonumber
\end{align}
respectively for $N_c^{eff}=2, 2.5$ and $3$. These interesting results suggest that the 
effective number of color might be taken greater than $2.5$ in the framework of 
the factorization hypothesis in case of $\Lambda_b$ decay.
It is worth comparing the theoretical branching ratio, $\mathcal{BR}^{th}(\Lambda_b 
\to \Lambda J/\psi)$,  with the experimental one~\cite{Eidelman:2004wy}, 
$\mathcal{BR}^{exp}(\Lambda_b \to \Lambda J/\psi)=(4.7\pm2.1 \pm 1.9)\times 10^{-4}$.
%
%
\section{Conclusion}\label{section6}
%
Calculations of the angular distributions as well as branching ratios 
of the process $\Lambda_b \to \Lambda V$ with $\Lambda \to P {\pi}^-$ 
and  $V \to {\ell}^+ {\ell}^-$ or $V \to h^+ h^-$ have been performed by 
using the helicity formalism and stressing on the correlations which arise 
among the final decay products. In all these calculations, particular role 
of the $\Lambda_b$ polarization has been put into evidence. The initial 
polarization, ${\cal P}_{\Lambda_b}$, appears  explicitly in the polar angle 
distribution of the $\Lambda$ hyperon in the $\Lambda_b$ rest-frame. 
Similarly,  the azimuthal angle distributions of both proton and ${\ell}^-$ 
in the $\Lambda$ and $V$ frames, respectively, depend directly on the 
$\Lambda_b$ polarization. Furthermore, a first computation of the asymmetry 
parameter, ${\alpha}_{As}$, in $\Lambda_b$ decays into $\Lambda V(1^-)$ has 
been performed as well as the longitudinal polarization of the vector meson, 
${\rho}_{00}^V$, which is shown to be dominant $(\ge 56\%)$.

On the other hand, it is well known that the violation of $CP$ symmetry 
via the CKM mechanism is one of the corner-stone of the Standard Model
of particle physics. Looking for TR violation effects in baryon decays 
provides us a new field of research: firstly as a complementary test of 
$CP$ violation by assuming the correctness of the $CPT$ theorem 
and, secondly, as a possibility to  search for processes beyond the Standard Model.
In particular, triple product correlations, which are $T$-odd under time 
reversal, can be extensively investigated in $\Lambda_b$ decays. However, 
this latter aim requires both experimental and theoretical improvements in 
order to increase our knowledge of $b$-physics.  

%
\subsection*{Acknowledgments}
The authors  are indebted to their colleagues of the LHCB Clermont-Ferrand team 
for internal discussions regarding this promising research subject of 
Time Reversal.  
\section*{Appendix}
\appendix
%
%
\section{Angular distributions}
%
%
\subsection{$\boldsymbol{\Lambda_b  \to  \Lambda V}$ decay}
%
Writing the hadronic matrix element, $\mathcal{M}_{\Lambda_b}(\lambda_1,\lambda_2)$,  
into two parameters according to the final helicity value such as,
\begin{eqnarray}
{|\mathcal{M}_{\Lambda_b} (\pm1/2)|}^2 = \   {|\mathcal{M}_{\Lambda_b}(\pm1/2,0)|}^2 
+ {|\mathcal{M}_{\Lambda_b}(\mp1/2,\mp1)|}^2\ ,
\end{eqnarray}
and by introducing the  helicity asymmetry parameter, ${\alpha}_{As}^{\Lambda_b}$, 
defined by,
\begin{eqnarray}
 \alpha_{AS}^{\Lambda_b} =  \frac{{|\mathcal{M}_{\Lambda_b}{(+1/2)}|}^2 -
 {|\mathcal{M}_{\Lambda_b}{(-1/2)}|}^2}{{|\mathcal{M}_{\Lambda_b}{(+1/2)}|}^2 +
 {|\mathcal{M}_{\Lambda_b}{(-1/2)}|}^2}\ ,
\end{eqnarray} 
the final angular distribution, $W{(\theta, \phi)}$, deduced\footnote{Integrating 
Eq.~(7) over the angles $\theta_1,\phi_1,\theta_2$ and $\phi_2$, and summing over the
helicities $\lambda_3, \lambda_5$ and $\lambda_6$.} from Eq.~(7) and  expressed as 
\begin{eqnarray}
W{(\theta, \phi)}  \propto    1 + {\alpha_{AS}^{\Lambda_b}}{\cal P}_{\Lambda_b}{\cos\theta} +
 2{\alpha_{AS}^{\Lambda_b}} \Re e \Big[ {\rho}_{ij}^{\Lambda_b}{\exp{(-i\phi)}} \Big]\sin{\theta}\ , 
\end{eqnarray} 
puts into evidence the parity violation.
%
\subsection{$\boldsymbol{\Lambda \to  P {\pi}^-}$ decay}
%
From Eq.~(7), integrating over the angles $\theta,\phi,\theta_2$ and  $\phi_2$ and 
summing over vector helicity states, the general formula for proton angular distributions, 
$W_1{(\theta_1,\phi_1)}$, in the $\Lambda$ frame reads as,
\begin{multline}
W_1{(\theta_1,\phi_1)}    \propto  \\
\frac12 \Biggl\{ ({\rho}_{ii}^{\Lambda}+{\rho}_{jj}^{\Lambda})+
({\rho}_{ii}^{\Lambda}-{\rho}_{jj}^{\Lambda}) 
\alpha_{AS}^{\Lambda} \cos\theta_1   
 -  \frac{\pi}{2}{\cal P}_{\Lambda_b}\alpha_{AS}^{\Lambda}
  \Re e \Big[{\rho}_{ij}^{\Lambda} \exp{(i\phi_1)} \Big] \sin{\theta_1}\Biggr\} \ ,
\end{multline} 
where the PDM elements, ${\rho}_{ij}^{\Lambda}$, of the baryon $\Lambda$ are (to a 
normalization factor):
\begin{align}
{\rho}^{\Lambda}_{ii} &=   {\int_{\theta_2,\phi_2}}G^V_{00}{(\theta_2, \phi_2)} 
\ {|\mathcal{M}_{\Lambda_b}{(\pm1/2,0)}|}^2 + 
{\int_{\theta_2,\phi_2}}G^V_{\pm 1 \pm1}{(\theta_2, \phi_2)} {|\mathcal{M}_{\Lambda_b}
{(\pm 1/2,\pm 1)}|}^2\ , \nonumber \\
{\rho}^{\Lambda}_{ij} &= {\int_{\theta_2,\phi_2}}
{G^V_{00}}{(\theta_2, \phi_2)}  
\mathcal{M}_{\Lambda_b}{(-1/2,0)} \mathcal{M}_{\Lambda_b}^*{(1/2,0)}\ .
\end{align}
The hermitian matrix, $G^V_{\lambda_2 {\lambda}^{\prime}_2}{(\theta_2,\phi_2)}$, 
describing the  process, $V \to  {\ell}^+ {\ell}^-$ or   $ V \to h^+ h^-$, has the 
following form:
\begin{eqnarray}
 G^V_{\lambda_2 {\lambda}^{\prime}_2}{(\theta_2,\phi_2)} = \sum_{\lambda_5,\lambda_6}
{|\mathcal{M}_{V}(\lambda_5,\lambda_6)|}^2
d^1_{\lambda_2 m_2}{(\theta_2)} d^1_{{\lambda}^{\prime}_2 m_2}{(\theta_2)} \exp{i(\lambda_2 
-{\lambda}^{\prime}_2)\phi_2}\ , 
\end{eqnarray}
with $m_2 =\lambda_5-\lambda_6$. In case of lepton pair in the final state, because of parity 
conservation, two hadronic matrix elements, $\mathcal{M}_{V}(\frac12,\pm \frac12)$, are 
necessary whereas only one, $\mathcal{M}_{V}(0,0)$, is required in case of pseudo-scalar mesons.
%
\subsection{$\boldsymbol{V \to  {\ell}^+ {\ell}^-  (h^+ h^-)}$}
%
Vector meson, V, decaying into a lepton pair or a hadronic one is described by the 
$(3 \times 3)$ hermitian matrix $G^V_{\lambda_i {\lambda}^{\prime}_i}{(\theta_i,\phi_i)}$. 
The angular distributions, $W_2{(\theta_2,\phi_2)}$, in the $V$ rest-frame, are obtained by 
integrating Eq.~(7) over the angles $\theta,\phi,\theta_1,\phi_1$ and summing over the two 
$\Lambda$ helicity states:
\begin{multline}
W_2{(\theta_2, \phi_2)}    \propto     
 ({\rho}_{ii}^{V}+{\rho}_{jj}^{V})(G^V_{00}(\theta_2,\phi_2)+G^V_{\pm 1 \pm1}(\theta_2,\phi_2))
\\
-  \frac{\pi}{4}{\cal P}_{\Lambda_b}  \Re e \Big[{\rho}_{ij}^V \exp{(i\phi_2)} \Big] \sin{2\theta_2}\ ,
\end{multline}
where  the PDM elements, ${\rho}_{ij}^{V}$, of the meson $V$ are (to a normalization factor):
\begin{align}
{\rho}^{V}_{ii} &=  {\int_{\theta_1,\phi_1}}  F^{\Lambda}_{\lambda_1 {\lambda}^{\prime}_1}{(\theta_1,\phi_1)}
\Biggl[  \delta_{\lambda_2 \lambda_2^{\prime}}  {|\mathcal{M}_{\Lambda_b}{(\pm1/2,0)}|}^2 + 
\delta_{\lambda_2 \pm\lambda_2^{\prime}}  {|\mathcal{M}_{\Lambda_b}{(\pm 1/2,\pm 1)}|}^2 \Biggr]\ , \nonumber \\
{\rho}^{V}_{ij} &= 
{\int_{\theta_1,\phi_1}}  F^{\Lambda}_{\lambda_1 {\lambda}^{\prime}_1}{(\theta_1,\phi_1)}
 \Biggl[\Bigl\{\mathcal{M}_{\Lambda_b}{(1/2,0)} \mathcal{M}_{\Lambda_b}^*{(1/2,1)} + h.c.\Bigr\}
\nonumber
\\ & \hspace{4.4cm} - \Bigl\{\mathcal{M}_{\Lambda_b}{(-1/2,0)} \mathcal{M}_{\Lambda_b}^*{(-1/2,-1)} 
+ h.c.\Bigr\}\Biggr]\mathcal{M}_{V \to hh(ll)} \ ,
\end{align}
where, $\mathcal{M}_{V\to hh(ll)}$, takes the following form according to the given decay:
\begin{align}
\mathcal{M}_{V \to hh(ll)}=&|\mathcal{M}_{V}(0,0)|^2\ , \;\; {\rm for} \;\;V \to h^+ h^-\ , \nonumber \\
\mathcal{M}_{V \to hh(ll)}=& |\mathcal{M}_{V}(1/2,-1/2)|^2 - 2 |\mathcal{M}_{V}(+1/2,+1/2)|^2\ , 
\;\; {\rm for} \;\; V \to l^+ l^-\ .
\nonumber
\end{align}
The function, $F^{\Lambda}_{\lambda_1 {\lambda}^{\prime}_1}{(\theta_1,\phi_1)}$, containing the 
decay dynamical part  of $\Lambda \to P \pi$ has the form,
\begin{multline}
F^{\Lambda}_{\lambda_1 {\lambda}^{\prime}_1}{(\theta_1,\phi_1)} = 
 \exp{i(\lambda_1-{\lambda}^{\prime}_1)\phi_1} \\
\Big({|\mathcal{M}_\Lambda(+1/2,0)|}^2 d^{1/2}_{\lambda_1 {1/2}}{(\theta_1)}
d^{1/2}_{{\lambda}^{\prime}_1 {1/2}}{(\theta_1)} +
{|\mathcal{M}_\Lambda(-1/2,0)|}^2 d^{1/2}_{\lambda_1 {-1/2}}{(\theta_1)}
d^{1/2}_{{\lambda}^{\prime}_1 {-1/2}}{(\theta_1)}\Big)\ ,
\end{multline}
where two hadronic matrix elements, $\mathcal{M}_\Lambda(\pm 1/2,0)$, are necessary 
to fully describe the intermediate resonance.
%
%

\begin{figure}[hpb]
\begin{center}
\includegraphics*[width=0.32\columnwidth]{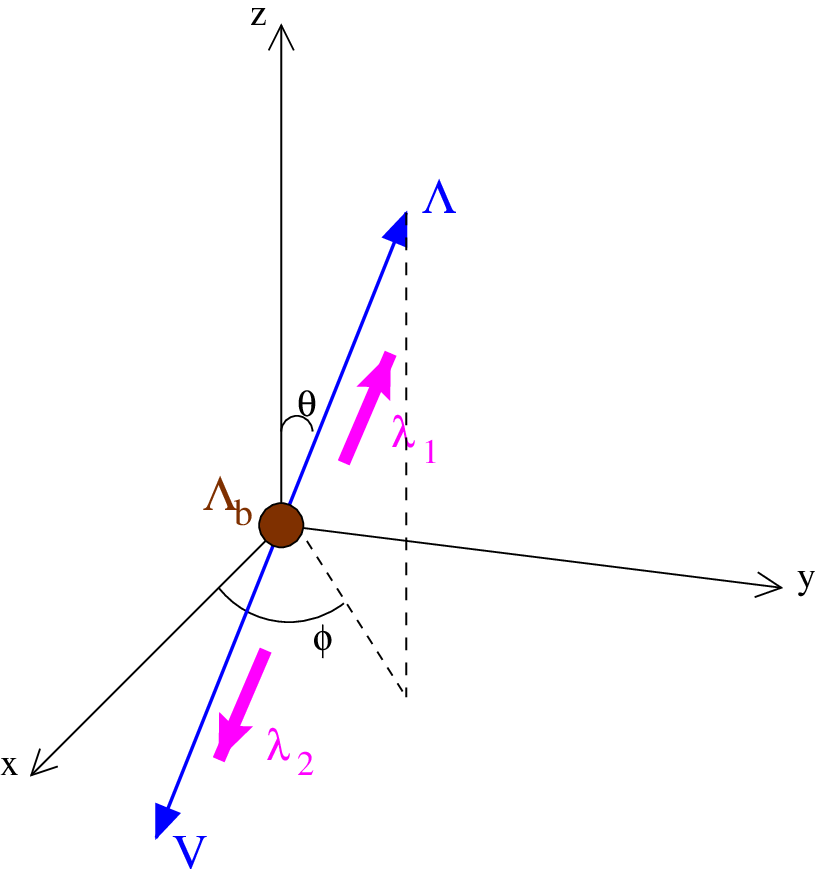}
\includegraphics*[width=0.32\columnwidth]{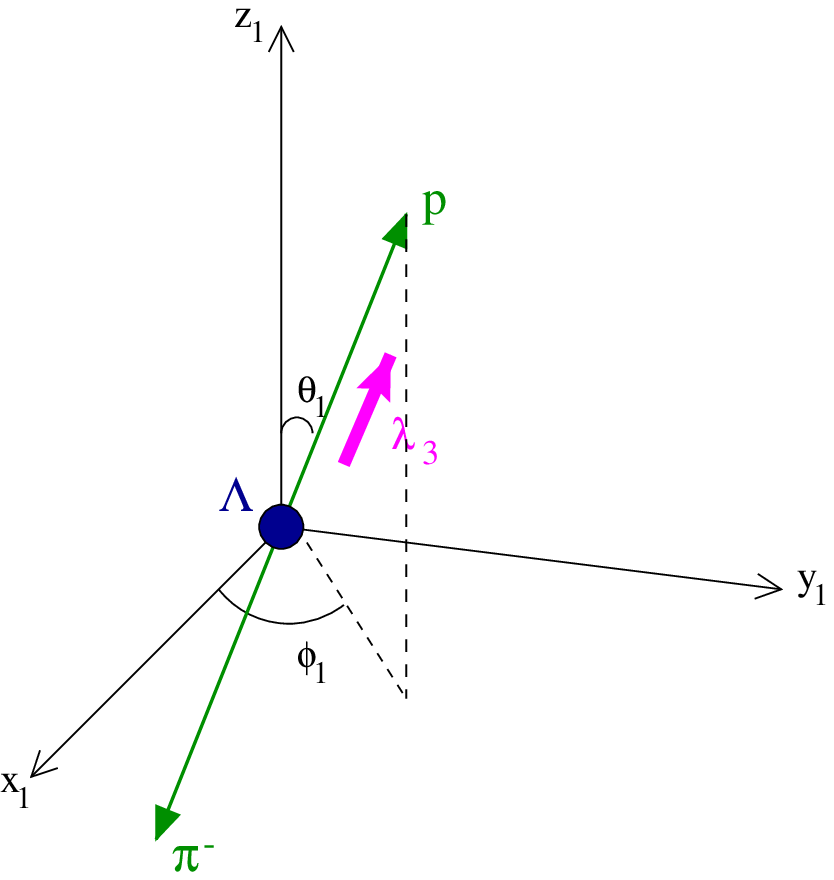}
\includegraphics*[width=0.32\columnwidth]{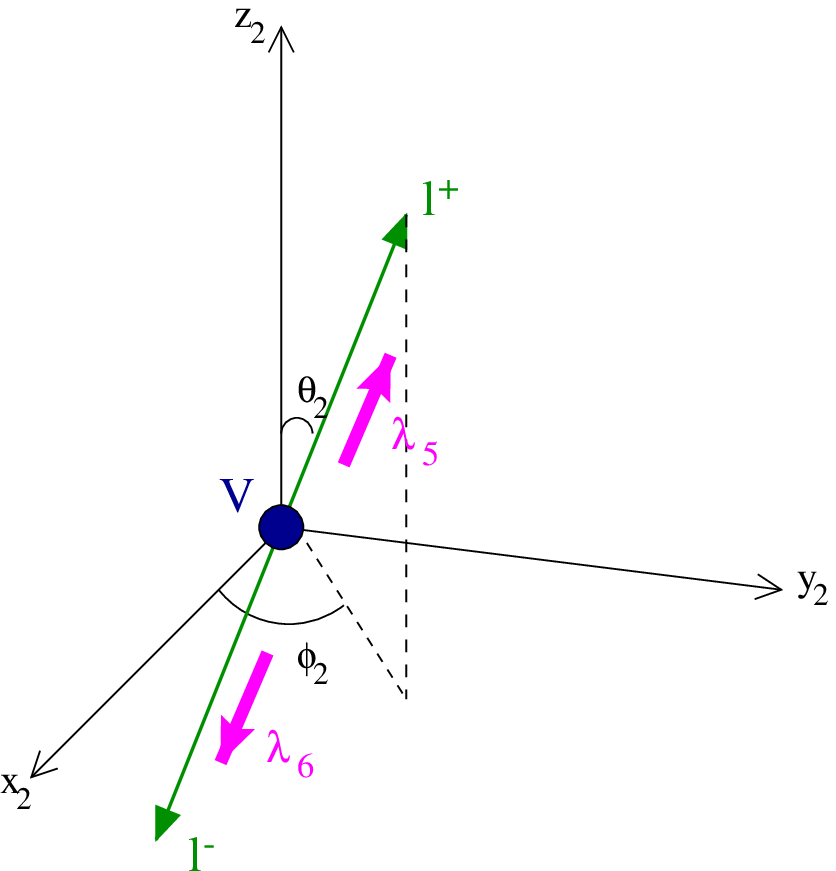}
\caption{Left-handed: $\Lambda_b$ decay in its transversity frame. Right-handed: 
helicity frames for the $\Lambda$ and vector meson V decays, respectively.}
\label{fig2}
\end{center}
\end{figure}
\begin{figure}
\begin{center}
\includegraphics*[width=0.325\columnwidth]{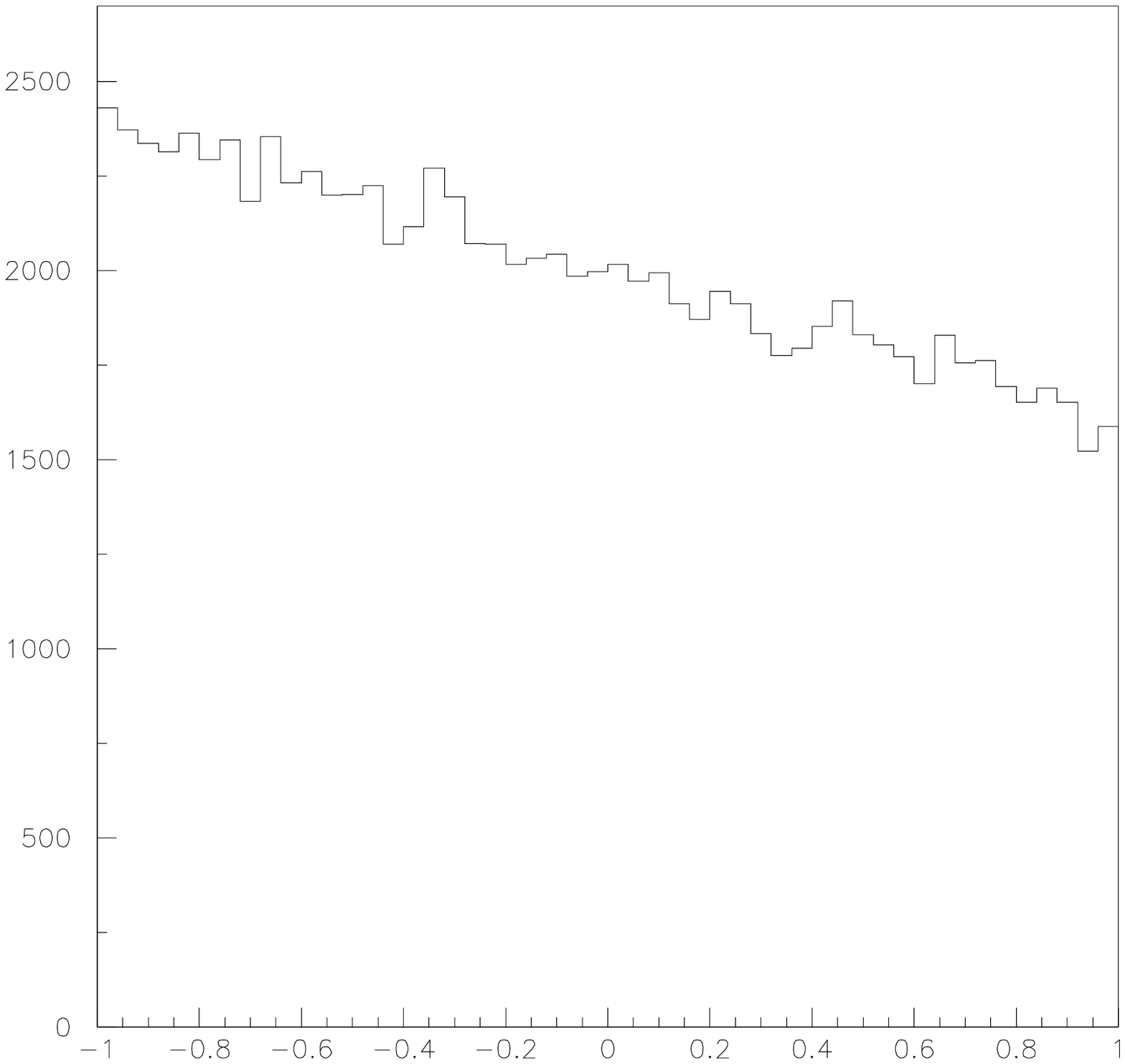}
\includegraphics*[width=0.325\columnwidth]{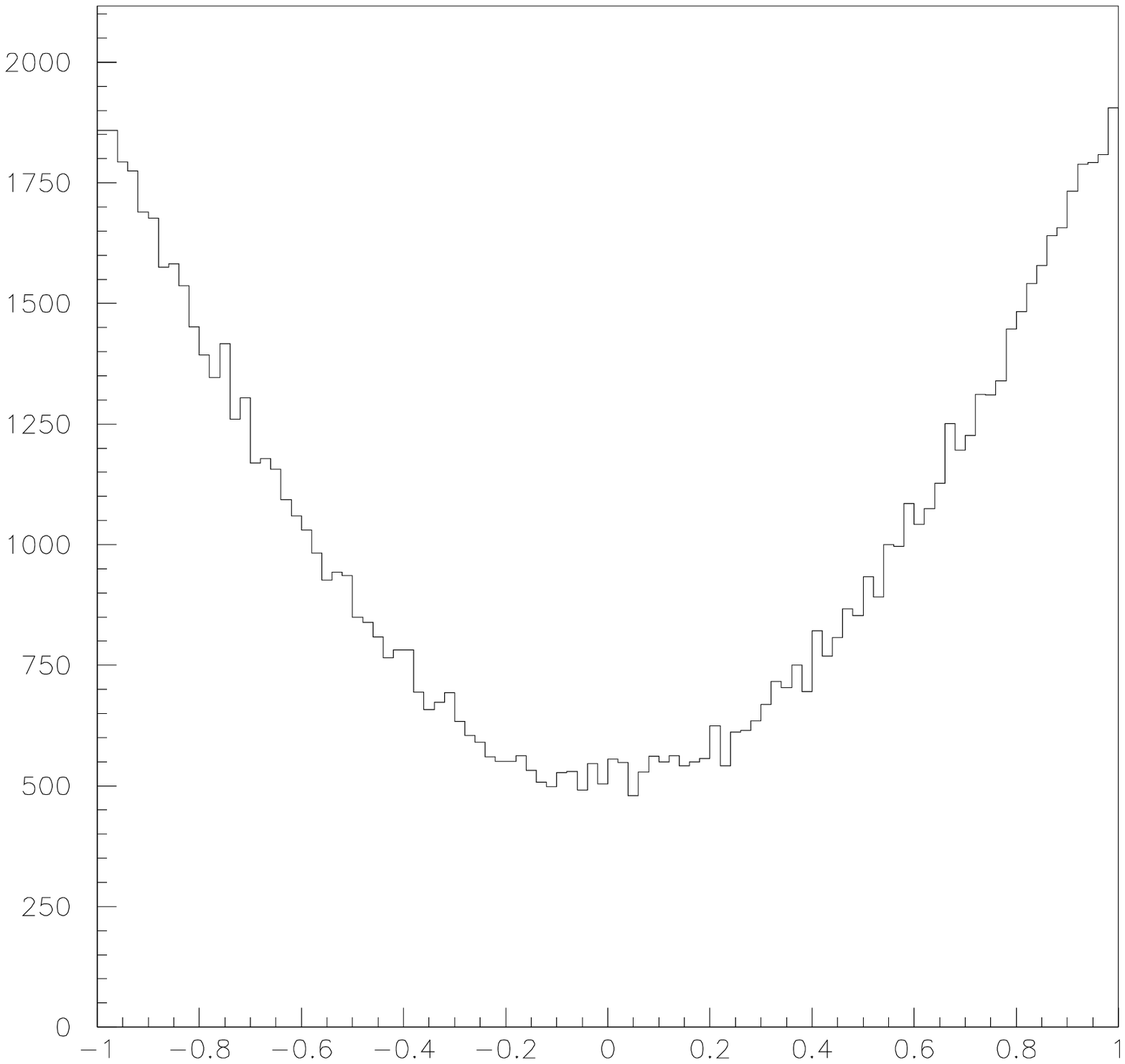}
\includegraphics*[width=0.325\columnwidth]{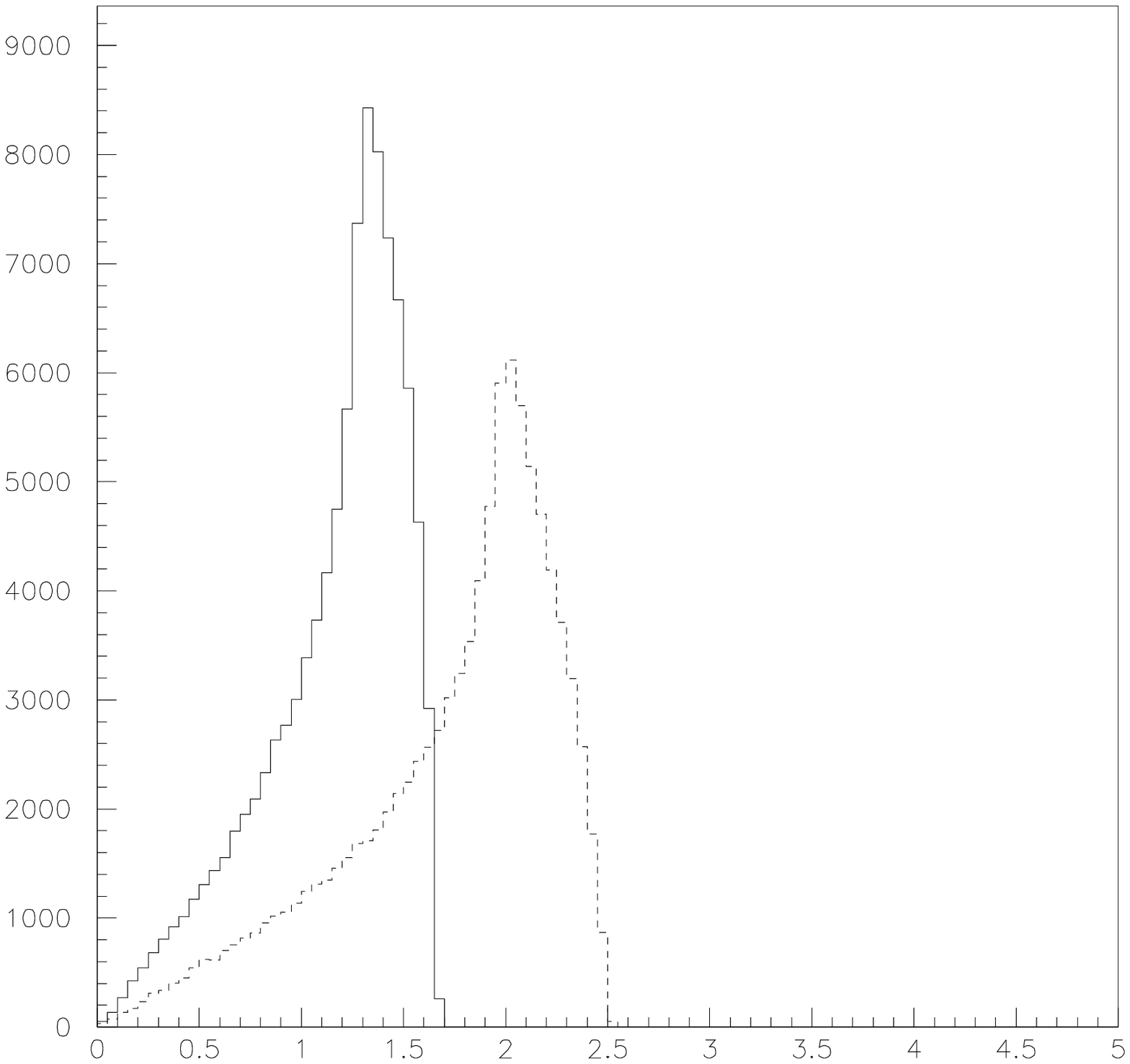}
\includegraphics*[width=0.325\columnwidth]{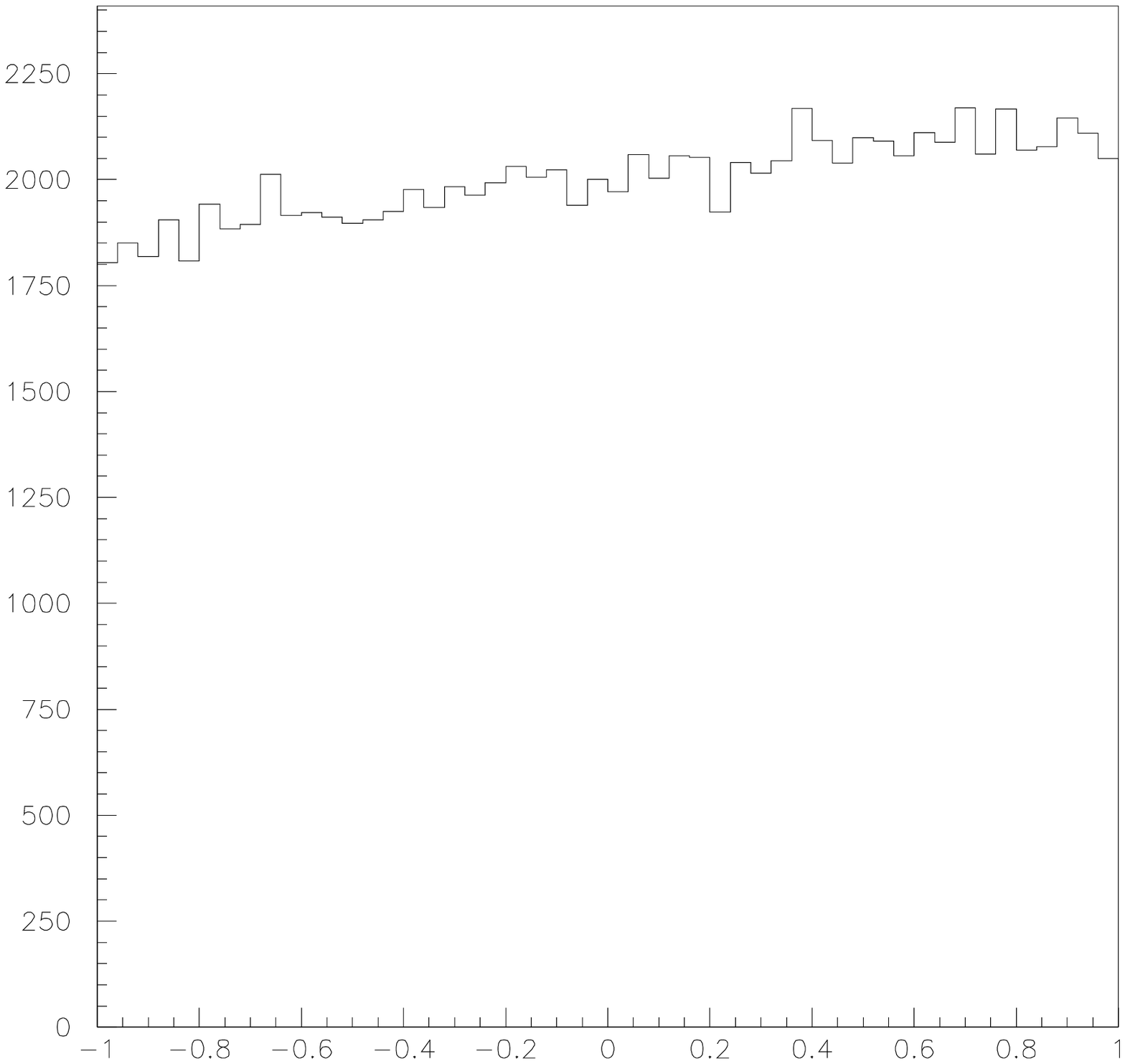}
\includegraphics*[width=0.325\columnwidth]{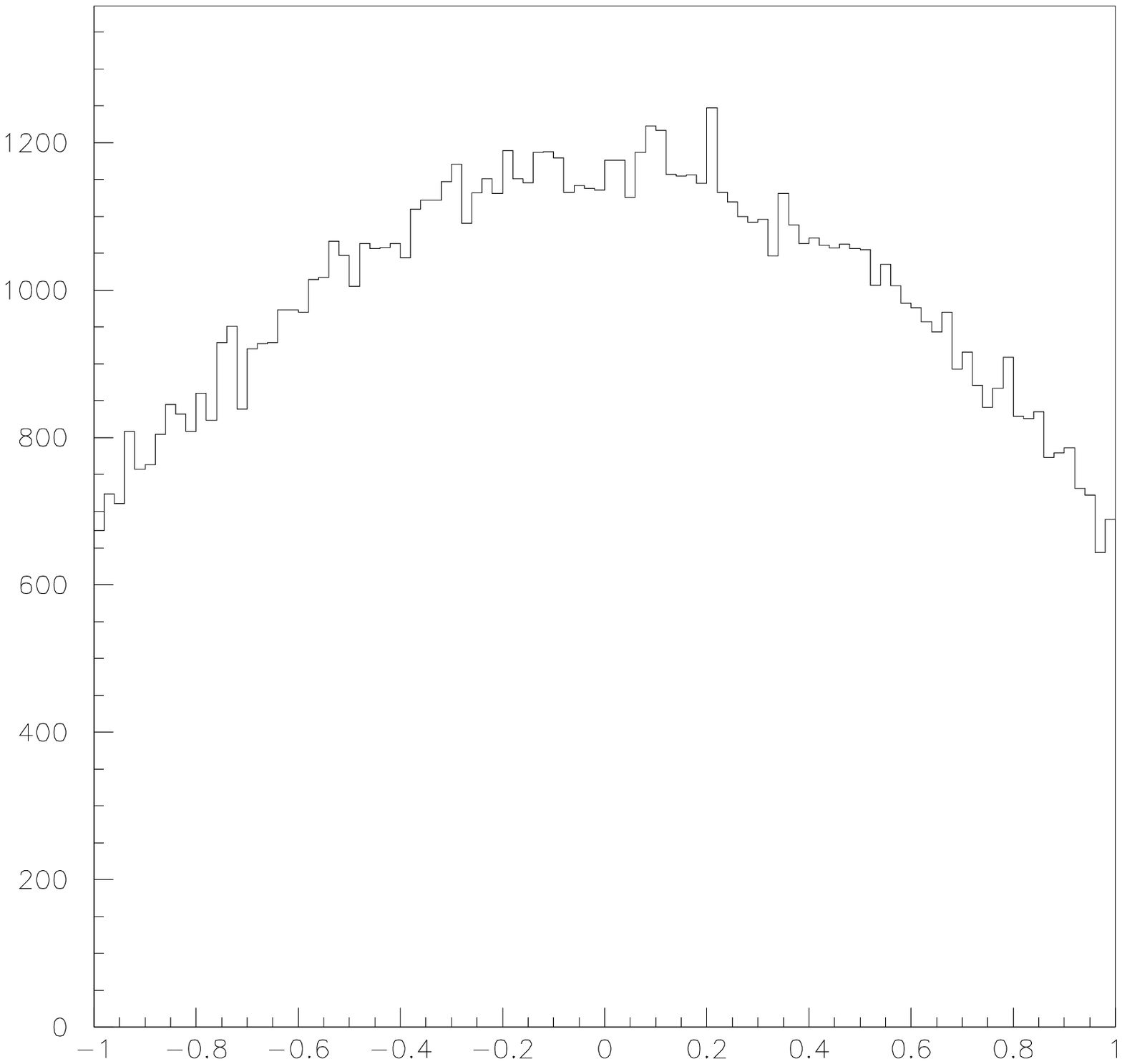}
\includegraphics*[width=0.325\columnwidth]{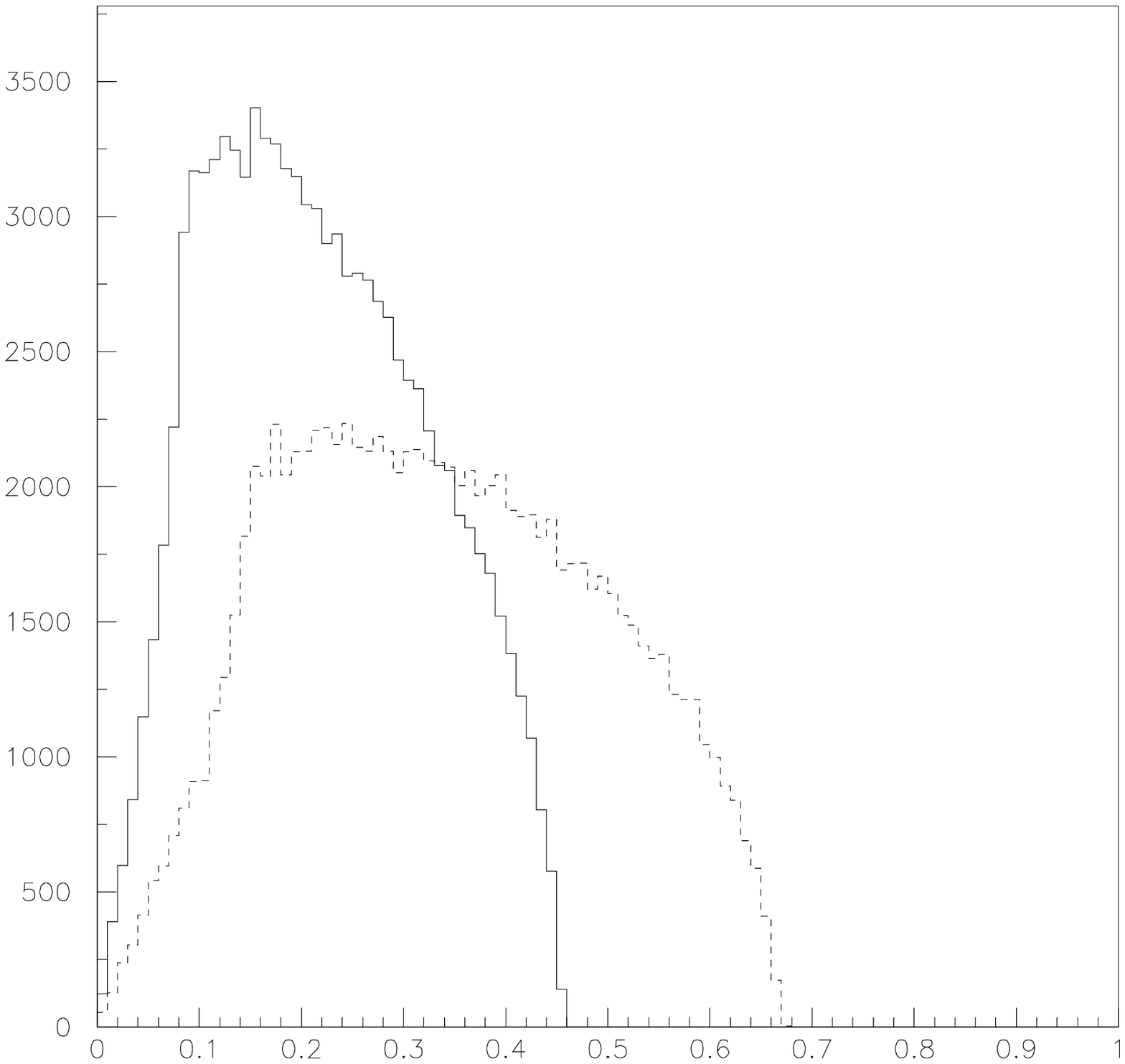}
\caption{First two columns: polar angular distributions in the intermediate resonance rest-frames 
for  the following values of $\mathcal{P}_{\Lambda}$ and ${\rho}_{00}^V$ parameters:
$(31\%, 65.5\%)$ in the case of $\Lambda_b \to \Lambda {\rho}^0$ (upper histograms: $\cos {\theta}_P$ 
(left side) and $\cos {\theta}_{{\pi}^-}$ (right side)), and $(-9\%, 55.5\%)$ in the case of
$\Lambda_b \to \Lambda J/{\psi}$ (lower histograms: $\cos {\theta}_P$ (left side) and 
$\cos {\theta}_{{\mu}^-}$ (right side)). \newline
Third column: proton and pion ($\Lambda$ daughters) transverse momentum, $P_{\perp}$, in the $\Lambda_b$ rest-frame, 
 in the case of $\Lambda {\rho}^0$ channel (dashed line) 
and $\Lambda J/{\psi}$ channel (full line), respectively. 
Upper histogram for  proton $P_{\perp}$-spectra and lower histogram for pion $P_{\perp}$-spectra.}
\end{center}
\end{figure}
\end{document}